# Pressure dependence of the low- temperature crystal structure and phase transition behaviour of CaFeAsF and SrFeAsF: A synchrotron x-ray diffraction study

S. K. Mishra[1], R. Mittal[1], S. L. Chaplot[1], S. V. Ovsyannikov[2], D. M. Trots[2], L. Dubrovinsky[2], Y. Su[3], Th. Brueckel[3,4], S. Matsuishi[5], H. Hosono[5], G. Garbarino[6]

[1]*Solid State Physics Division, Bhabha Atomic Research Centre, Trombay, Mumbai 400 085, India*
[2]*Bayerisches Geoinstitut, University Bayreuth, Universitatsstrasse 30, D-95440 Bayreuth, Germany*
[3]*Juelich Centre for Neutron Science JCNS, Forschungszentrum Juelich GmbH, Outstation at FRM II, Lichtenbergstr. 1, D-85747 Garching, Germany*
[4]*Juelich Centre for Neutron Science JCNS and Peter Grünberg Institut PGI, JARA-FIT, Forschungszentrum Juelich GmbH, D-52425 Juelich, Germany*
[5]*Frontier Research Center, Tokyo Institute of Technology, 4259 Nagatsuta-cho, Midori-ku,Yokohama 226-8503, Japan*
[6]*European Synchrotron Radiation Facility, BP 220, 38043 Grenoble, France.*

We report systematic investigation of high pressure crystal structures and structural phase transition upto 46 GPa in CaFeAsF and 40 GPa in SrFeAsF at 40 K using powder synchrotron x-ray diffraction experiments and Rietveld analysis of the diffraction data. We find that CaFeAsF undergoes orthorhombic to monoclinic phase transition at $P_c$ = 13.7 GPa while increasing pressure. SrFeAsF exhibits coexistence of orthorhombic and monoclinic phases over a large pressure range from 9 to 39 GPa. The coexistence of the two phases indicates that the transition is of first order in nature. Unlike in the 122 compounds ($BaFe_2As_2$ & $CaFe_2As_2$) we do not find any collapse tetragonal transition. The transition to a lower symmetry phase (orthorhombic to monoclinic) in 1111 compounds under pressure is in contrast with the transition to a high symmetry phase (orthorhombic to tetragonal) in 122 type compounds. On heating from 40 K at high pressure, CaFeAsF undergoes monoclinic to tetragonal phase transition around 25 GPa and 200 K. Further, it does not show any post-tetragonal phase transition and remains in the tetragonal phase upto 25 GPa at 300 K. The $dP_c/dT$ is found to be positive for the CaFeAsF & $CaFe_2As_2$, however the same was not found in case of $BaFe_2As_2$. We discuss observations of structural evolution in the context of superconductivity in these and other Fe-based compounds. It appears that the closeness of the Fe-As-Fe bond angle to its ideal tetrahedral value of $109.47^0$ might be associated with occurrence of superconductivity at low temperature.

# I. Introduction

Crystal structures and phase transitions are vital to superconductivity in iron arsenide based compound [1-20]. These compounds occur in five different structural classes [2] (namely: FeSe, LiFeAs, BaFe$_2$As$_2$, LaFeAsO/SrFeAsF and Sr$_3$Sc$_2$O$_5$Fe$_2$As$_2$) but share a common layered structure based on a planar layer of iron atoms joined by tetrahedrally coordinated pnictogen (Pn=P, As) or chalcogen (S, Se, Te) anions arranged in a stacked sequence separated by alkali, alkaline-earth or rare-earth (Ba, Ca, Sr, Eu) and oxygen/fluorine blocking layers. It is now widely thought that the interaction that leads to the high-Tc superconductivity originates within these common iron layers, similar in nature to the common copper oxygen building block found in the copper oxide (cuprate). However structurally three key differences are found among the FeAs and cuprate compounds. First, in the FeAs compounds the pnictogen/chalcogen anions are placed above and below the planar iron layer as opposed to the planar copper-oxygen structure of the cuprates. Second ability to substitute or dope directly in to the active pairing layers of FeAs compounds and finally the parent compounds of the new Fe based superconductors share a similar electronic structure with all five d orbitals of the Fe contributing to a low density of states at the Fermi level which is in contrast to the cuprates where parent compounds are Mott insulators with well-defined local magnetic moments. These similarities have inspired a flurry of theoretical and experimental works [1-30] in Fe pnictides based materials.

At ambient condition, these compounds crystallize in tetragonal symmetry with no magnetic order (i.e paramagnetic in nature). The parent compounds of iron-pnictides undergo a first or second order structural transition below room temperature (typically be in the range of 100-210 K), from tetragonal to orthorhombic structure, and magnetic transition from non-magnetic to stripe antiferro magnetic structure [2-7, 22, 24, 29, 30]. The structural transition and magnetic ordering can be happen simultaneously or successively depending on the compound. It has been confirmed both experimentally and theoretically that the magnetic order of Fe at low temperature is stripe-like antiferromagnetism often referred to as spin density wave (SDW) [2-7]. Upon changing the carrier concentration, applying external pressure or by charge neutral doping, the magnetic order suppresses and the materials become superconducting. In the 1111-family (both RFeAsO and MFeAsF, R=rare earth and M=Ca and Sr), magnetic transition temperature ($T_N$) is lower than structural transition temperature ($T_S$). This seems to suggest that the magnetic transition is induced by the structural transition. In case of M′Fe$_2$As$_2$ (M′=Ba, Sr, Ca, and Eu) compounds, (at ambient pressure with decreasing temperature) the structural and magnetic transitions are found [29-30] to happen simultaneously (a first-order transition). However, recent high pressure powder x-ray synchrotron diffraction studies on BaFe$_2$As$_2$ show that at low temperature (40 K), these transitions occur at different pressures [19]. It is not clear whether the magnetic transition is induced by the structural transition and what is the driving force of the structural transition. These two important questions are crucial to understand the formation of the stripe antiferromagnetic order in the parent compounds.

The magnetism is intimately related to the crystal structure, both in terms of the Fe-As bond length and As-Fe-As bond angle. The antiferromagnetic ordering can be suppressed by various ways like: by changing the carrier concentration, applying external pressure or by charge neutral doping. The common FeAs building block is considered a critical component to stabilize superconductivity. The combination of strong bonding between Fe-Fe and Fe-As sites (and even interlayer As-As in the 122-type systems), and the geometry of the $FeAs_4$ tetrahedra plays a crucial role in determining the electronic and magnetic properties of these compounds. For instance, the two As-Fe-As tetrahedral bond angles seem to play a crucial role in optimizing the superconducting transition temperature. The highest $T_c$ values are found only when an As-Fe-As/Fe-As-Fe tetrahedral bond angle is close to the ideal value of 109.47 degree. The detailed interplay between the crystal structure, magnetic ordering, and superconductivity is hardly understood.

High pressure experiments play an important role in the field of superconductivity and also provide information about the understanding of its mechanism. The superconducting transition temperature ($T_c$) in FeAs compounds is found to increase [14-21] over 50 K by application of pressure. Despite the importance of the evolution of superconducting transition temperature ($T_c$) with pressure to understand the mechanism of superconducting properties of Fe based materials, there is lack of information on the detailed pressure dependence of their structural properties. Recently, we reported the pressure effects on $CaFe_2As_2$ and $BaFe_2As_2$ (122 type compounds) using powder synchrotron x-ray diffraction technique. Rietveld analysis of the high pressure powder x-ray diffraction data showed [19] that at 300 K in the Ba-compound the collapsed tetragonal transition occurs at 27 GPa. The transition pressure value is found to be much higher as compared to the Ca-compound where the transition occurs at 1.7 GPa. However, at low temperature (33K), structural phase transition from the orthorhombic to tetragonal phase in the Ba-compound occurred at about 29 GPa (while increasing pressure), which is much higher than the transition pressure of 0.3 GPa at 40 K as known in case of the Ca-compound [19]. We have not found any evidence of a post collapsed tetragonal phase transition in $CaFe_2As_2$ up to 51 GPa (at 300 K) and 37.8 GPa (at 40 K). It is important to note that transition to a collapsed phase occurs in the two compounds at nearly the same values of unit cell volume and $c_t/a_t$ ratio. Although five different types [2] of Fe based superconductor have been reported in literature.The 122 FeAs based superconductors seems to be the most studied of the five types. On the other hand 1111 type compound have not got considerable attention of scientific community. In this paper, we present systematic investigation of pressure effect on crystal structure and structural phase transition behaviour of CaFeAsF and SrFeAsF at 40 K using powder diffraction technique. Detail Rietveld analysis of the diffraction data shows that both the compounds undergo a structural phase transition from orthorhombic to monoclinic phase on compression. While CaFeAsF undergoes a fairly sharp transition at around 13.7 GPa, SrFeAsF exhibits a large phase coexistence region (from 9 to 40 GPa) with the orthorhombic phase fraction continuously decreasing with increase of pressure. Possible correlation between structure evolutions with pressure and superconductivity in Fe-based superconductor is also discussed.

## II. Experiments

The pressure dependent powder x-ray diffraction measurements were carried out using the ID-27 beam line at the European Synchrotron Radiation Facility (ESRF, Grenoble, France). An applied pressure was generated by membrane diamond anvil cells (DACs). We employed a stainless steel gasket pre-indented to the thickness of ~ 40-50 μm, with a central hole of 150 μm in diameter and filled with helium as pressure transmitting media. The pressure was determined from the shift of the fluorescence line of the ruby .A powdery sample of ~30-40 μm in diameter and 10 μm in thickness was situated in the center of one of the diamond anvils tips. The wavelength of the x-ray (0.3738 Å) was selected and determined using a Si(111) monochromator and the iodine K-edge. The sample to image plate (MAR345) detector distance was refined using the standard diffraction data of Si. To cooling the DAC, a continuous helium flow CF1200 DEG Oxford cryostat was used. Precaution was taken to obtain stable temperature and pressure conditions prior to each acquisition. The precision and accuracy of the temperature measurement is better than 0.1 K and 0.2 K, respectively. During the measurements, the CaFeAsF and SrFeAsF samples were first cooled to 40 K, and then pressure was increased along a path indicated in Fig. 1. Typical exposure times of 20 seconds were employed for the measurements.

The 2-dimensional powder images were integrated using the program FIT2D [31] to yield intensity *vs.* 2θ plot. The structural refinements were performed using the Rietveld refinement program FULLPROF [32]. In all the refinements, the background was defined by a sixth order polynomial in 2θ. A Thompson-Cox-Hastings pseudo-Voigt with axial divergence asymmetry function was chosen to define the profile shape for the powder synchrotron x-ray diffraction peaks. The scale factor, background, half-width parameters along with mixing parameters, lattice parameters and positional coordinates, were refined.

## III. Result and Discussion

The powder synchrotron x-ray diffraction measurements for MFeAsF (M=Ca, Sr) at ambient conditions confirmed single-phase samples consistent with the published reports [22, 24-26]. The effects of pressure inhomogineity on the phase transition behaviour of FeAs based compounds have been reported in literature [14-19]. In the present measurements, we have used helium as pressure transmitting medium, which is believed [33] to give the best hydrostatic conditions at present. However, effect of inhomogeneity could not be completely ruled out and could have some influence on the results obtained on these compounds.

## (A) Phase transition of CaFeAsF at 40 K

At ambient condition, CaFeAsF crystallizes in tetragonal structure with space group *P4/nmm* without magnetic ordering. On cooling, it undergoes a tetragonal to orthorhombic phase transition at 134 K while magnetic ordering in the form of a spin-density wave sets in at around 114 K respectively. Figures 2 (a) and (b) show a Rietveld refinement of powder synchrotron data at (0.6 GPa & 300 K) and (5.8 GPa & 40 K) respectively. Splitting of (220) peak of tetragonal phase (at ambient temperature) unambiguously confirms the orthorhombic structure at 40 K (see inset).

As a function of pressure the powder diffraction patterns show significant changes especially in terms of dissimilar broadening of various peaks. The strongest changes have been observed in peaks around $2\theta=11°$ which become broader with increasing pressure above 12 GPa. Detail Rietveld refinement of the powder diffraction data shows that diffraction patterns at 40 K could be indexed using the orthorhombic structure (space group *Cmma*) up to 12 GPa (Fig. 3). It is evident from Fig. 3 (b) that the diffraction data at 20.3 GPa cannot be indexed with the orthorhombic phase. Extra broadening (splitting) of peaks suggests lowering of the symmetry. We have explored various possibilities to identify the correct space group and found that monoclinic structure with space group *P12/n1* (c ba setting of *P112/n*) could successfully index all the peaks (see Fig. 3(b)). This structure is already well documented [5, 6] in literature and many iso-structural FeAs-based materials like LaFeAsO [6], undergo tetragonal to monoclinic phase transition with temperature. A careful inspection and analysis of diffraction data reveal that CaFeAsF transforms to the monoclinic structure at $P_c$= 13.7 GPa even though the monoclinic distortion is quite small. The monoclinic angle β shows very small change from 90.5° to 91.2° on increase in pressure from 13.7 GPa to 46.2 GPa. However, the small monoclinicity has significantly improved the fit quality between the observed and calculated profiles as shown in Figure 3. The detailed structural parameters and goodness of fit for CaFeAsF at selected pressure and 40 K, as obtained from powder synchrotron diffraction data, are given in Table I. Schematic diagrams of the crystal structure in the orthorhombic and monoclinic phases are shown in Figure 4.

## (B) Phase transition of SrFeAsF at 40 K

SrFeAsF also crystallizes in tetragonal structure with space group *P4/nmm* at ambient condition. On cooling, it undergoes a structural phase transition to orthorhombic phase around 180 K, followed by paramagnetic to anti-ferromagnetic transition at 133 K. Figure 5 (a) and (b) depict results of Rietveld refinement of powder synchrotron data at (0.6 GPa & 300 K) and (5.8 GPa & 40 K) respectively. At low temperature, we found that refinement for SrFeAsF can be carried out in the orthorhombic structure consistent with literature. The observation is unambiguously confirmed by comparing the $(220)_T$ peak of tetragonal phase at ambient temperature (see inset).

As in the case of CaFeAsF, the powder diffraction patterns of SrFeAsF also show significant changes with pressure. To obtain the pressure dependence of the structural parameters, detailed Rietveld refinement of the powder synchrotron x-ray diffraction data are carried out. Similar to CaFeAsF, we notice broadening/splitting of some of the peaks around 9 GPa. At this pressure, the diffraction data could not be refined using either the orthorhombic or the monoclinic phase as indicated in Fig. 6. However, a two-phase refinement with both the orthorhombic and monoclinic space groups is found successful and all the observed diffraction peaks could be indexed (see Fig. 6).

We find that orthorhombic and monoclinic phases coexist over a large pressure range from 9 GPa to the highest pressure of 40 GPa attained in our experiment. The percentage of the monoclinic phase continuously increases with pressure and reaches 98 % at 40 GPa. The coexistence of both the phases indicates that the structural phase transition from orthorhombic to monoclinic phase is of first order. The fit between the observed and calculated profiles is quite satisfactory and some of them are shown in Figure 6. The detailed structural parameters and goodness of fit for SrFeAsF at selected pressure and 40 K, as obtained from powder synchrotron diffraction data, are given in Table II.

In our earlier measurements for $BaFe_2As_2$ we found that the coexisting region for orthorhombic and tetragonal phase was very large [19]. In case of the fluorine based 1111 type Ca/Sr compounds the transition pressures are found to be very similar. These observations are in contrast to the case of $BaFe_2As_2$ and $CaFe_2As_2$, where the compound with smaller ionic radii ($CaFe_2As_2$) show phase transition at much lower pressure (0.3 GPa) at 40 K as compared to the $BaFe_2As_2$, where transition at 40 K occur at 29 GPa [19].

### (C) Pressure evolution of structural parameters at 40 K

As mentioned above, superconductivity in iron arsenide materials is associated with lattice distortion and suppression of magnetic ordering. The detailed interplay between the crystal structure, magnetic ordering, and superconductivity is hardly understood up to now which is to some extent due to the lack of precise structural data. At ambient conditions, MFeAsF (M= Ca/Sr) crystallizes in the ZrCuSiAs type structure (space group *P4/nmm*, Z = 2). The crystal structures of all the iron-pnictides share a common two-dimensional FeAs layer, where Fe atoms form a 2D square sublattice with As atoms sit at the center of these square, but off the Fe plane (above and below the plane alternately). These are made of alternating Ca/SrF and FeAs layers. The Fe and F atoms lie in planes, while the As and Ca/Sr atoms are distributed on each side of these planes following a chessboard pattern. They undergo a tetragonal to orthorhombic phase transition at 134, 180 K followed by magnetic transitions at 114, 133K for CaFeAsF and SrFeAsF respectively [22, 24]. Applications of internal pressure (chemical pressure) suppress both orthorhombic structure and antiferromagnetic state, leading to emergence of superconductivity. For example, the critical superconducting transition temperature $T_C$ ~4K in the Co-substituted at Fe site in SrFeAsF, $T_C$ ~22 K in the Co- substituted at Fe site in CaFeAsF, and $T_C$ ~ 31 K, 52 K, 56 K respectively in the La, Nd and Sm substituted at Sr site in SrFeAsF were observed [23-28]. In addition to this, it was also found that systematically replacing R from La, Ce, Pr, Nd and Sm in $RFeAsO_{1-\delta}$ resulted in a gradual decreases in the a- axis lattice parameters and an increase in superconducting transition temperature. Superconducting transition temperature ($T_c$) of different Fe-based superconductor is indeed correlated to their structural properties. A systematic trend between Tc and the Fe-As-Fe angles would be expected to be found, because the exchange couplings are directly related to the Fe-As-Fe bond angle and Fe-Fe/Fe-As bond distances. It is found that the highest $T_c$ is obtained when the Fe-As-Fe reaches the ideal value of $109.47^o$ for the perfect FeAs tetrahedron with the least lattice distortion. This suggests that the most effective way to increase $T_c$ in Fe-based superconductors is to decrease the deviation of the Fe-As-Fe bond angle from the ideal FeAs tetrahedron, as the geometry of the FeAs tetrahedron might be correlated with the density of states near the Fermi energy.

In view of this, we have carried out detailed structural analysis as a function of pressure. Figure 7 and 8 show the pressure dependence of the structural parameters of CaFeAsF and SrFeAsF at 40 K. For easy comparison, orthorhombic lattice parameters (a, b, and c) are converted using the relation $a_o = a/\sqrt{2}$, $b_o = b/\sqrt{2}$ and $c_o = c$. It is clear from Figs. 7 and 8 that at 40 K on increasing pressure, the lattice parameters monotonically decrease in the entire range of our measurements for CaFeAsF, while the a lattice parameter of SrFeAsF exhibits anomalous increase beyond 30 GPa. The pressure variation of $FeAs_4$ polyhedral volume, As-Fe-As bond angle and Fe-As and Fe-Fe bond lengths obtained from the Rietveld analysis of diffraction data are shown in Figure 9 and 10. It is clear from these figures that the polyhedral volume decreases with increasing pressure. In case of the CaFeAsF, the Fe-Fe and Fe-As bond lengths reveal anomaly around 11.6 GPa. However, for SrFeAsF, Fe-Fe bond lengths show anomalous increase beyond 30 GPa. A similar anomalous increase was also observed in $BaFe_2As_2$ [18] that was associated with loss of magnetism. The angle between As-Fe-As is quite different from the ideal tetrahedral angle of $109.47^0$ and shows anomalous behaviour with pressure. It is also evident from Figs. 7-10 that the nature of phase transition seems to be different in CaFeAsF and SrFeAsF. As stated earlier, the transition is sharp in CaFeAsF, whereas SrFeAsF exhibits coexistence of the two phases over a large range of pressure.

Temperature dependence of the electrical resistivity measurements for CaFeAsF at different pressures were carried out by Okada *et al.* [20, 21] using the piston-cylinder-type cell and the cubic anvil press. They found that the magnetic transition is suppressed by pressure above 5 GPa, the resistivity anomaly disappears, and superconductivity is observed. Further, the resistivity loss becomes larger and the superconducting transition shifts to lower temperature with increasing pressure up to 20 GPa. In our experiment at a fixed temperature of 40 K with increasing pressure, the Fe-As-Fe bond angle decreases over 5 to 10 GPa and then increases with further increasing pressure. A clear correlation of the bond-angle and superconductivity is difficult to establish in view of the different pressure-temperature paths followed in the two experiments. To the best of our knowledge, the high pressure resistivity measurements are still missing for SrFeAsF.

In order to determine the bulk modulus $B$ at zero pressure and its pressure derivative $B'$, the pressure-volume data were fitted by a third order Birch-Murnaghan equation. For CaFeAsF at 40 K, the fit give values of $B$ =82.1 ± 1.3 GPa and $B'$ =5.9 ± 1.8 for the orthorhombic phase, while the monoclinic phase is found to be less compressible with $B$ =135.2 ± 3.4 GPa, and $B'$ =3.2 ± 0.2  The $B$ and $B'$ values extracted from the pressure volume relation in the orthorhombic phase (40 K, 5.38-31 GPa) of SrFeAsF are 111.1 ± 5.3 GPa and 4.0 ± 0.5 and in the monoclinic phases (40 K, 11.26-39 GPa) the corresponding values are 114.2 ± 7.3 GPa and 4.0 ± 0.6,  respectively. The fitted ambient pressure volumes for the orthorhombic and monoclinic phases of CaFeAsF at 40 K are $V_o$= 255.45 ± 0.3 Å$^3$ and 123.00 ± 0.66 Å$^3$ respectively. However $V_o$ for the orthorhombic and monoclinic phases for SrFeAsF at 40 K are $V_o$= 275.55 ± 0.3 Å$^3$ and 134.06 ± 0.86 Å$^3$ respectively.

## (D) Phase transition of CaFeAsF on heating at high pressure

Due to experimental limitations, we could not increase pressure on CaFeAsF beyond 46 GPa at 40 K. However, to see the effect of temperature on the newly stabilized monoclinic phase in CaFeAsF (at 40 K and around 40 GPa), we carried out measurements at different conditions (temperature-pressure) as shown in figure 1. Careful analysis of the diffraction data show that there is an abrupt change in the diffraction pattern at 25 GPa and 200 K when the monoclinic splitting/broadening in Bragg reflections around at $11.5^0$ and $16.5^0$ disappears (insets of Figure 11).  Detailed Rietveld analyses of the diffraction data reveal that the sample undergoes a structural phase transition from monoclinic to tetragonal phase at around 25 GPa and 200 K and remains in the tetragonal phase at 25 GPa and 300 K. The fit between the observed and calculated profiles is shown in Figure 11. For SrFeAsF, we could not able to measure any data point while decreasing the pressure and increasing of temperature.  Using a third order Birch-Murnaghan equation, the computed values of the bulk modulus $B$ and its pressure derivative $B'$ for CaFeAsF at room temperature are 107.7 ± 3.5 GPa and $B'$ =2.5 ± 0.3 respectively.

We have summarized our observations of various phases in CaFeAsF and SrFeAsF with pressure and temperature in Figure 1. It is interesting to note that at 40 K, CaFeAsF shows orthorhombic to monoclinic phase transition at $P_c$ =13.7 GPa, whereas at room temperature tetragonal phase is stable upto 25 GPa. The observation of $dP_c/dT>0$ is similar to that observed in CaFe$_2$As$_2$ where collapsed tetragonal phase transition occurs at a lower pressure (0.3 GPa) at a low temperature (50 K) in comparision to 1.7 GPa at 300 K. However, this behaviour of dPc/dT may not hold for Ba/Sr compounds [19]. In addition to this, the transition to a lower symmetry phase (orthorhombic to monoclinic) in 1111 (CaFeAsF/SrFeAsF) compound under pressure is in contrast with the high symmetry phase (orthorhombic to tetragonal) in 122 (BaFe$_2$As$_2$/ CaFe$_2$As$_2$ ) type compounds.

## IV. Conclusion

We have investigated the effect of pressure on the crystal structure and structural phase transition behaviour at 40 K in CaFeAsF and SrFeAsF using powder synchrotron x-ray diffraction and Rietveld analysis technique. We found that both the compounds undergo structural phase transition from orthorhombic to monoclinic phase with increasing pressure. CaFeAsF undergoes a fairly sharp orthorhombic to monoclinic phase transition at 13.7 GPa with increasing pressure. On the other hand, SrFeAsF exhibits coexistence of orthorhombic and monoclinic phases over a large pressure range from 9 to 39 GPa. The coexistence of the two phases indicates that the transition is of first order. On heating from 40 K at high pressure, CaFeAsF undergoes monoclinic to tetragonal phase transition around 25 GPa and 200 K. Further, it does not show any post tetragonal phase transition up to 25 GPa at 300 K. We note that the 1111-compound (CaFeAsF) undergoes phase transition to a lower symmetry phase (i. e., orthorhombic to monoclinic) under pressure in contrast with the transition to a higher symmetry phase (i. e., orthorhombic to tetragonal) observed in 122-type compounds (BaFe$_2$As$_2$ and CaFe$_2$As$_2$). We have also determined the bulk modulii in these compounds that confirm their soft nature analogous to other compounds in the FeAs family.

**Acknowledgments**
R. Mittal and S. K. Mishra thank Department of Science and Technology (DST), India for providing financial support to carry out synchrotron x-ray diffraction at European Synchrotron Radiation Facility, Grenoble, France.

1. Y Izyumov, E Kurmaev, *High- Tc Superconductors Based on FeAs Compounds* (Spinger, Verlag, Hamburg 2010); J. S. Schilling, in Handbook of High temperature Superconductivity: Theory and experiments, edited by J. R. Schrieffer and J. S. Brooks (Spinger Verlag, Hamburg, 2007); H. Takashi and N. Mori, in Studies of High Temperature Superconductors, edited by A. Narlikar (Nova Science, New York, 1996).
2. J. Paglion and R L Greene, Nat. Phys. **6**, 645 (2010).
3. I. I. Mazin and M. D. Johannes, Nat. Phys. **5**, 141 (2009); G. Yu, Y. Li, E. M. Motoyama and M. Greven, Nat. Phys. **5**, 873 (2009).
4. Fa Wang and Dung-Hai Lee, Science **332**, 200 (2011).
5. J Zhao, Q. Huang, C D. Cruz, S. Li, J.W. Lynn, Y. Chen, M. A. Green, G. F. Chen, G. Li, Z. Li, J. L. Luo, N. L. Wang and P. Dai, Nature Materials **7** 953 (2008).
6. C de la Cruz, Q. Huang, J. W. Lynn, J. Li, W. Ratcliff II, J. L. Zarestky, H. A. Mook, G. F. Chen, J. L. Luo, N. L. Wang and P. Dai1, Nature (London) **453**, 899 (2008).
7. I. I. Mazin, D J. Singh, M. D. Johannes and M. H. Du, Phys. Rev. Lett. **101**, 057003 (2008); T. Yildirim, Phys. Rev. Lett. **101**, 057010 (2008); C. Feng, H. Yao, W.F. Tsai, J.P. Hu, and S A Kivelson, Phys. Rev. B **77,** 224509 (2009).
8. Y. Kamihara, T. Watanabe, M. Hirano and H. Hosono, J. Am. Chem. Soc. **130**, 3296 (2008).
9. K. Ishida, Y. Nakaii and H. Hosono, J. Phys. Soc. Jpn **78,** 062001 (2009); T. Imai, K. Ahilan, F. L. Ning, T. M. McQueen, and R. J. Cava Phys. Rev. Lett. **102**, 177005 (2009).
10. X. H. Chen, T. Wu, G. Wu, R. H. Liu, H. Chen, and D.F.Fang, Nature (London) **453**, 761 (2008).
11. M. Rotter, M. Tegel, and D. Johrendt, Phys. Rev. Lett. **101**, 107006(2008).
12. R. Mittal, L. Pintschovius, D. Lamago, R. Heid, K.-P. Bohnen, D. Reznik, S. L. Chaplot, Y. Su, N. Kumar, S. K. Dhar, A. Thamizhavel and Th. Brueckel, Phys. Rev. Lett. **102**, 217001 (2009).
13. R. Mittal, Y. Su, S. Rols, M. Tegel, S. L. Chaplot, H. Schober, T. Chatterji, D. Johrendt and Th. Brueckel, Phys. Rev. B **78**, 224518 (2008).
14. H. Takahashi, K. Igawa, K. Arii, Y. Kamihara, M. Hirano, and H. Hosono, Nature (London) **453**, 376 (2008).
15. M. S. Torikachvili, S. L. Bud'ko, N. Ni, and P. C. Canfield, Phys. Rev. Lett. **101**, 057006 (2008).
16. T. Park, E. Park, H. Lee, T.Klimczuk, E.D. Bauer, F. Ronning, and J. D. Thompson, J. Phys.: Condens. Matter **20**, 322204 (2008); P. L. Alireza, Y. T. Chris Ko, J. Gillett, C. M. Petrone, J. M. Cole, G. G. Lonzarich, and S. E. Sebastian, J. Phys.: Condens.: Matter **21**, 012208 (2009).

17. C. F. Miclea, M. Nicklas, H. S. Jeevan, D. Kasinathan, Z. Hossain, H. Rosner, P. Gegenwart, C. Geibel, and F. Steglich, Phys. Rev. B **79**, 212509 (2009); T. Terashima, M. Kimata, H. Satsukawa, A. Harada, K. Hazama, S. Uji, H. S. Suzuki, T. Matsumoto, and K. Murata, J. Phys. Soc. Jpn. **78**, 083701 (2009).
18. H. Okada, K. Igawa, H. Takahashi, Y. Kamihara, M. Hirano, H. Hosono, K. Matsubayashi, and Y. Uwatoko, J. Phys. Soc. Jpn.**77**, 113712 (2008).
19. R. Mittal, S. K. Mishra, S. L. Chaplot, S. V. Ovsyannikov, E. Greenberg, D. M. Trots, L. Dubrovinsky, Y. Su, Th. Brueckel, S. Matsuishi, H. Hosono, and G. Garbarino, Phys. Rev B, **83**, 054503 (2011).
20. H. Okada, H. Takahashi, S. Matsuishi, M. Hirano, and H. Hosono, K. Matsubayashi, Y. Uwatoko and H. Takahashi, Phys. Rev. B, **81**, 054507 (2010).
21. H. Okada, H. Takahashi, H. Takahashi, S. Matsuishi, M. Hirano, and H. Hosono, J. of Phys.:Confer. Series 200 012151(2010).
22. Y. Xiao, Y. Su, R. Mittal, T. Chatterji, T. Hansen, C. M. N. Kumar, S. Matsuishi, H. Hososno and Th. Brueckel, Phys. Rev. B, **79** 060504 (R) (2009).
23. Taku Hanna, Yoshinori Muraba, Satoru Matsuishi, Katsuaki Kodama, Shin-ichi Shamoto, Hideo Hosono, arXiv:1103.1177.
24. Y. Xiao, Y. Su, R. Mittal, T. Chatterji, T. Hansen, S. Price, C. M. N. Kumar, J. Persson, S. Matsuishi, Y. Inoue, H. Hososno and Th. Brueckel, Phys. Rev. B, **81** 094523 (2010).
25. S. Matsuishi, Y. Inoue, T. Nomura, H. Yanagi, M. Hirano, and H. Hosono, J. Am. Chem. Soc. **130**, 14428 (2008).
26. T. Nomura, Y. Inoue, S. Matsuishi, M. Hirano, J. E. Kim, K. Kato, M. Takata, and H. Hosono, Supercond. Sci. Technol. **22**, 055016 (2009).
27. G. Wu, Y. L. Xie, H. Chen, M. Zhong, R. H. Liu, B. C. Shi, Q. J. Li, X. F. Wang, T. Wu, Y. J. Yan, J. J. Ying, X. H. Chen, J. of Phys: Conden. Matter **21**, 142203(2009); Li-Fang Zhu, Bang-Gui Liu; Euro Phys. Lett. **85**, 67009 (2009).
28. X. Zhu, F. Han, P. Cheng, G. Mu, B. Shen, Hai-Hu Wen, Euro phys. Lett. **85**, 17011 (2009); Peng Cheng, Bing Shen, Gang Mu, Xiyu Zhu, Fei Han, Bin Zeng, Hai-Hu Wen, Europhys. Lett. **85**, 67003 (2009).
29. S. D. Wilson, Z. Yamani, C. R. Rotundu, B. Freelon, E. Bourret-Courchesne, and R. J. Birgeneau, Phys. Rev. B **79,** 184519 (2009); J.-Q. Yan, A. Kreyssig, S. Nandi, N. Ni, S. L. Bud'ko, A. Kracher, R. J. McQueeney, R. W. McCallum, T. A. Lograsso, A. I. Goldman, and P. C. Canfield, Phys. Rev. B **78**, 024516 (2008).
30. A. I. Goldman, D. N. Argyriou, B. Ouladdiaf, T. Chatterji, A. Kreyssig, S. Nandi, N. Ni, S. L. Bud'ko, P. C. Canfield, and R. J. McQueeney, Phys. Rev. B 78, 100506(R) (2008); Y. Xiao, Y.

Su, M. Meven, R. Mittal, C. M. N. Kumar, T. Chatterji, S. Price, J. Persson, N. Kumar, S. K. Dhar, A. Thamizhavel, and Th. Brueckel, Phys. Rev. B 80, 174424 (2009).

**31.** A. P. Hammersley, Report No. EXP/AH/95-01 (1995).

**32.** J. Rodriguez-Carvajal, Physica B **192**, 55 (1992).

**33.** N. Tateiwa and Y. Hag, Rev. Sci. Instrum. **80**, 123901 (2009).

**Table I:** Results of Rietveld refinement of the crystal structure for CaFeAsF at selected pressure and 40 K. The measurements were carried out using a focused x-ray monochromatic beam of wavelength=0.3738 Å. The data collected up to 25° have been used to determine the reported parameters.

| At 5.72 GPa & 40 K | | | | 29.3 GPa & 40 K | | |
|---|---|---|---|---|---|---|
| Orthorhombic phase (Space group: *Cmma*) | | | | Monoclinic phase (Space group: *P2/n*) | | |
| Atoms | Positional coordinates | | | Positional Coordinates | | |
| | X | Y | Z | X | Y | Z |
| Ca | 0 | 1/4 | 0.1640(1) | ¾ | 0.1750(2) | ¼ |
| Fe | 1/4 | 0 | 0.5000 | ¾ | 0.4892(1) | 3/4 |
| As | 0 | 1/4 | 0.6618(6) | ¾ | 0.6639(3) | ¼ |
| F | 1/4 | 0 | 0.0000 | ¾ | -0.0493(4) | 3/4 |
| **Lattice parameters** | | | | | | |
| a (Å) | | | 5.4012(1) | | | 3.7156(2) |
| b (Å) | | | 5.3733(2) | | | 7.5810(2) |
| c (Å) | | | 8.2897(3) | | | 3.6733(2) |
| | | | | | | β= 90.83 (3) degree |
| $R_p$=14.5; $R_{wp}$=25.7; $R_{exp}$=13.34; $\chi^2$= 3.71 | | | | $R_p$=17.8; $R_{wp}$=23.4; $R_{exp}$=13.39; $\chi^2$= 3.05 | | |

**Table II:** Results of Rietveld refinement of the crystal structure for SrFeAsF at selected pressure and 40 K. The measurements were carried out using a focused x-ray monochromatic beam of wavelength=0.3738 Å. The data collected up to 25° have been used to determine the reported parameters.

| At 5.8 GPa & 40 K | | | | 22.1 GPa & 40 K | | |
|---|---|---|---|---|---|---|
| Orthorhombic Phase | | | | Two phase: Orthorhombic+ Monoclinic | | |
| Space group: *Cmma* | | | | Space group: ( *Cmma*+ *P2/n*) | | |
| Atoms | Positional Coordinates | | | Positional Coordinates | | |
| | X | Y | Z | X | Y | Z |
| Sr | 0 | 1/4 | 0.1570(1) | 0/0.75 | 0.25/ 0.1455(3) | 0.1707(2)/ 0.25 |
| Fe | 1/4 | 0 | 0.5000 | 0.25/0.75 | 0/ 0.5131(1) | 0.5/0.75 |
| As | 0 | 1/4 | 0.6602(6) | 0/0.75 | 0.25/ 0.6703(2) | 0.6606(5)/0.25 |
| F | 1/4 | 0 | 0.0000 | 0.25/0.75 | 0/0.001(1) | 0/ 0.75 |
| **Lattice parameters** | | | | | | |
| a (Å) | | | 5.5125(3) | | | 5.34022(2)/3.7993(4) |
| b (Å) | | | 5.5380(4) | | | 5.3716(2)/8.1790(2) |
| c (Å) | | | 8.6680(2) | | | 8.2119(6)/3.7651(3) |
| | | | | | | β= 90.43 (5) degree |
| | | | | | | Phase fraction (%): 65(ortho.)/35 (mono.) |
| $R_p$=13.3; $R_{wp}$=21.4; $R_{exp}$=12.43; $\chi^2$= 2.96 | | | | $R_p$=10.0; $R_{wp}$=20.5; $R_{exp}$=12.01; $\chi^2$= 2.91 | | |

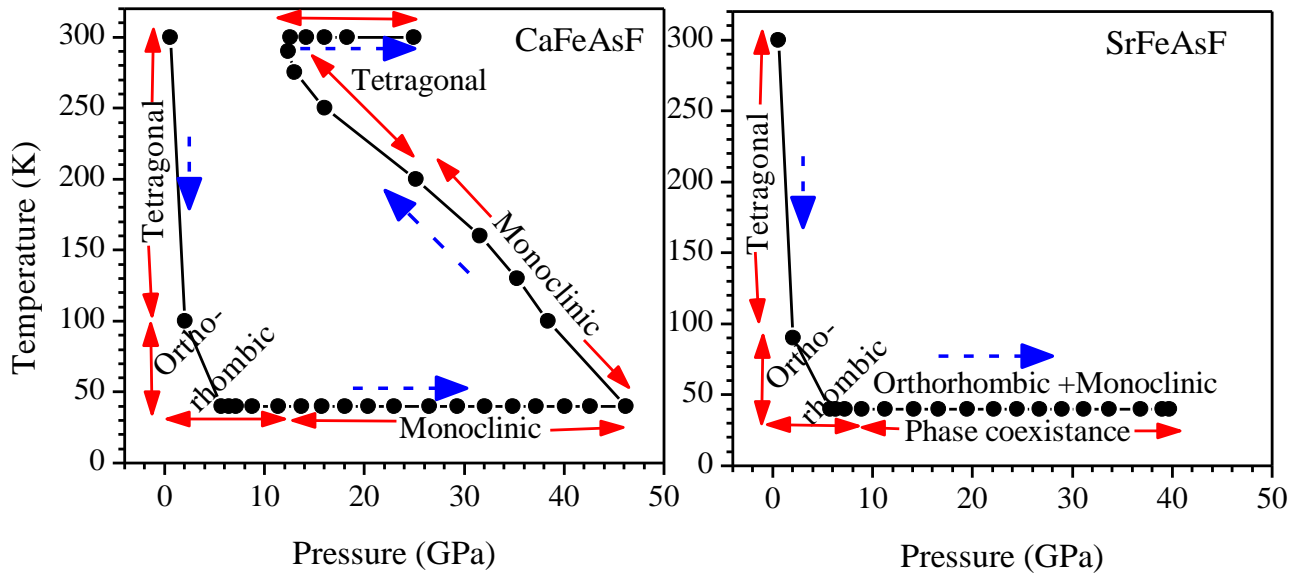

**Figure 1.** (Color online) The pressure-temperature path (indicated by blue dash line and arrows) as followed in our measurements on CaFeASF and SrFeAsF. Symbols correspond to the pressure-temperature conditions where measurements were made. The solid lines through the symbols are guide to the eye. The phases, namely, tetragonal, orthorhombic and monoclinic, as identified by Rietveld refinement of the diffraction data are indicated over ranges shown by red lines and arrows.

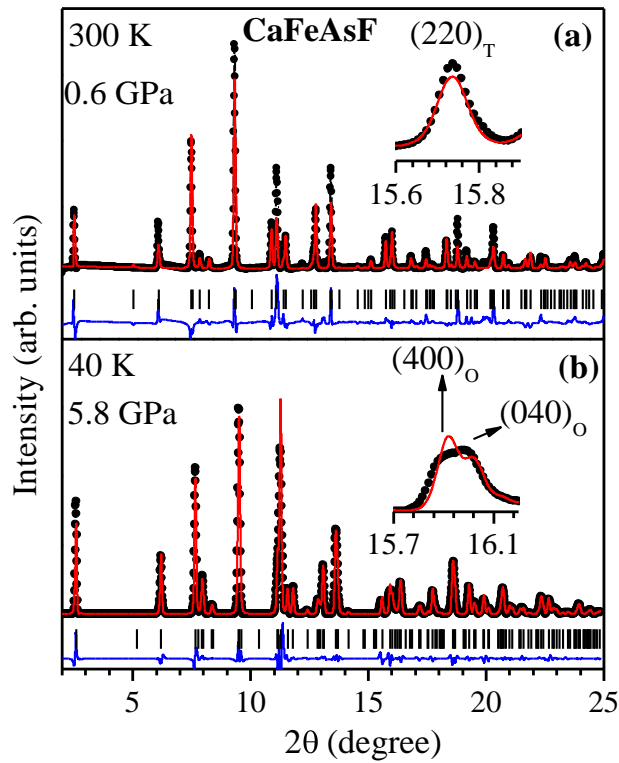

**Figure 2.** (Color online) Observed (black solid circle), calculated (continuous red line), and difference (bottom blue line) profiles obtained after the Rietveld refinement of CaFeAsF at (a) 0.6 GPa & 300 K in tetragonal phase (space group *P4/nmm*) and (b) 5.8 GPa & 40 K, in orthorhombic phase (space group *Cmma*). Inset in (a) show the (220) reflection of the tetragonal phase and in (b) show the spitting/broadening of the (220) reflection of the tetragonal phase at 40K and provides unambiguous signature for orthorhombic structure.

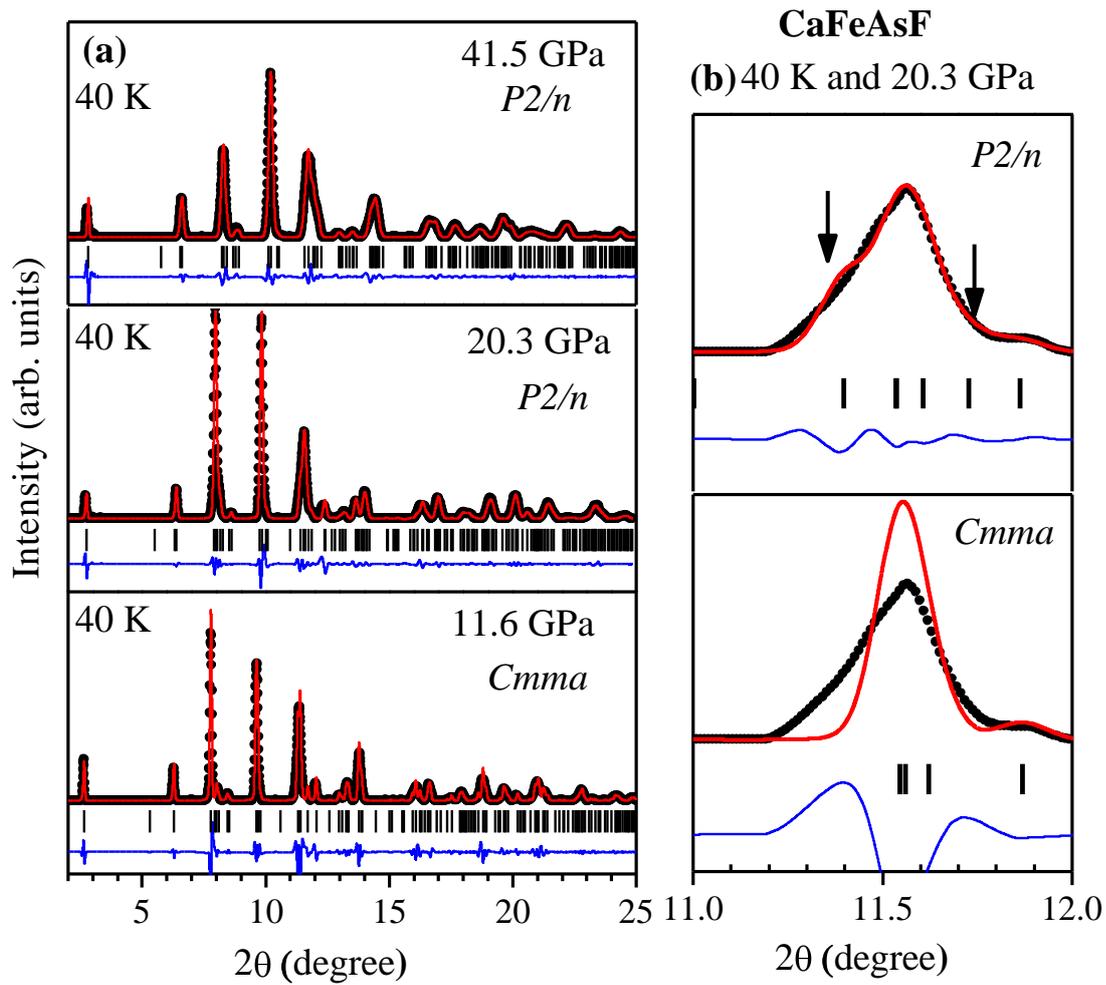

**Figure 3.** (Color online) (a) Observed (solid black circle), calculated (continuous red line), and difference (bottom blue line) profiles obtained after the Rietveld refinement of CaFeAsF at selected pressures and 40 K. The diffraction profiles at 11.6 GPa is refined using an orthorhombic (*Cmma*), while the profiles at 20.3, 29.3 and 41.5 GPa are refined monoclinic (*P2/n*) phases. (b) The refinement of the diffraction pattern at 20.3 GPa and 40 K with an orthorhombic phase (space group *Cmma*), and monoclinic (*P2/n*) phases.

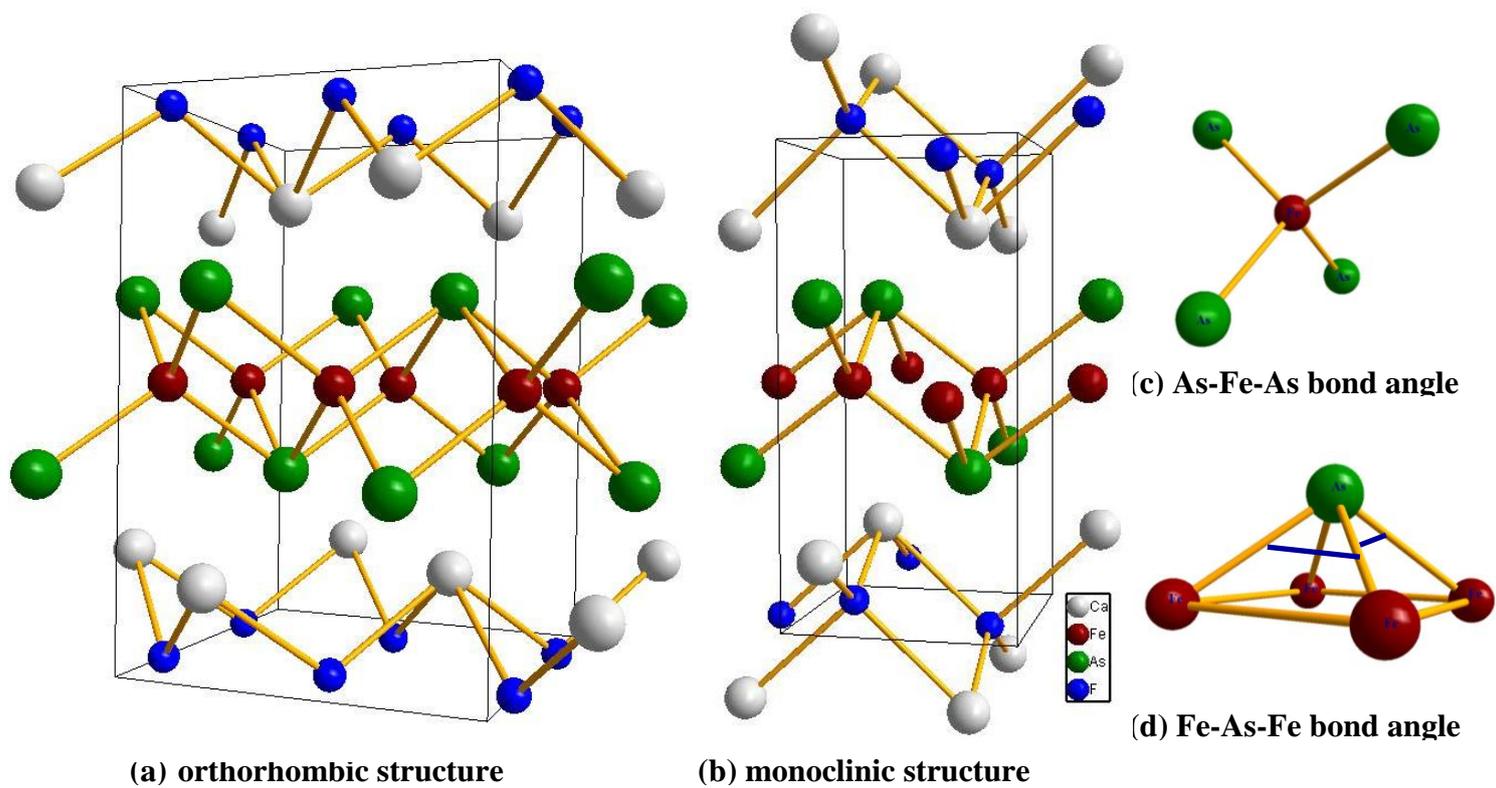

**Figure 4.** Schematic diagram of (a) orthorhombic phase (at 5.8 GPa & 40 K), (b) monoclinic phase (at 13.7 GPa & 40 K), and geometry of the FeAs$_4$ /AsF$_4$ units and definition of the (c) As-Fe-As and (d) Fe-As-Fe bond angles.

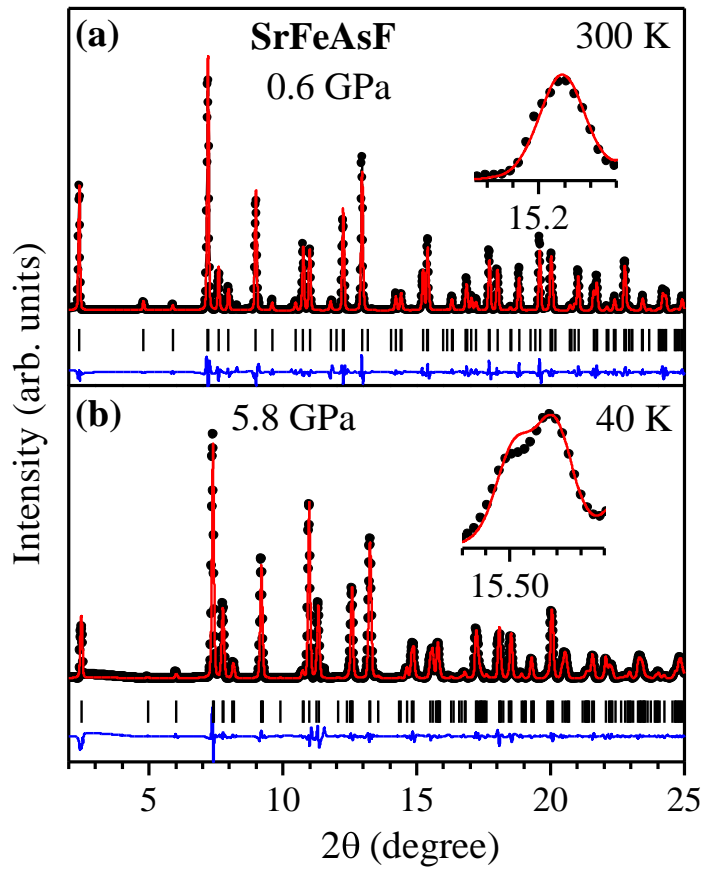

**Figure 5.** (Color online) Observed (solid black circle), calculated (continuous red line), and difference (bottom blue line) profiles obtained after the Rietveld refinement of SrFeAsF at (a) 0.6 GPa & 300K, in tetragonal phase (space group *P4/nmm*) and (b) 5.8 GPa & 40 K, in orthorhombic phase (space group *Cmma*). Inset in (a) show the (220) reflection of the tetragonal phase and in (b) show the spitting/broadening of the (220) reflection of the tetragonal phase at 40K and provides unambiguous signature for orthorhombic structure.

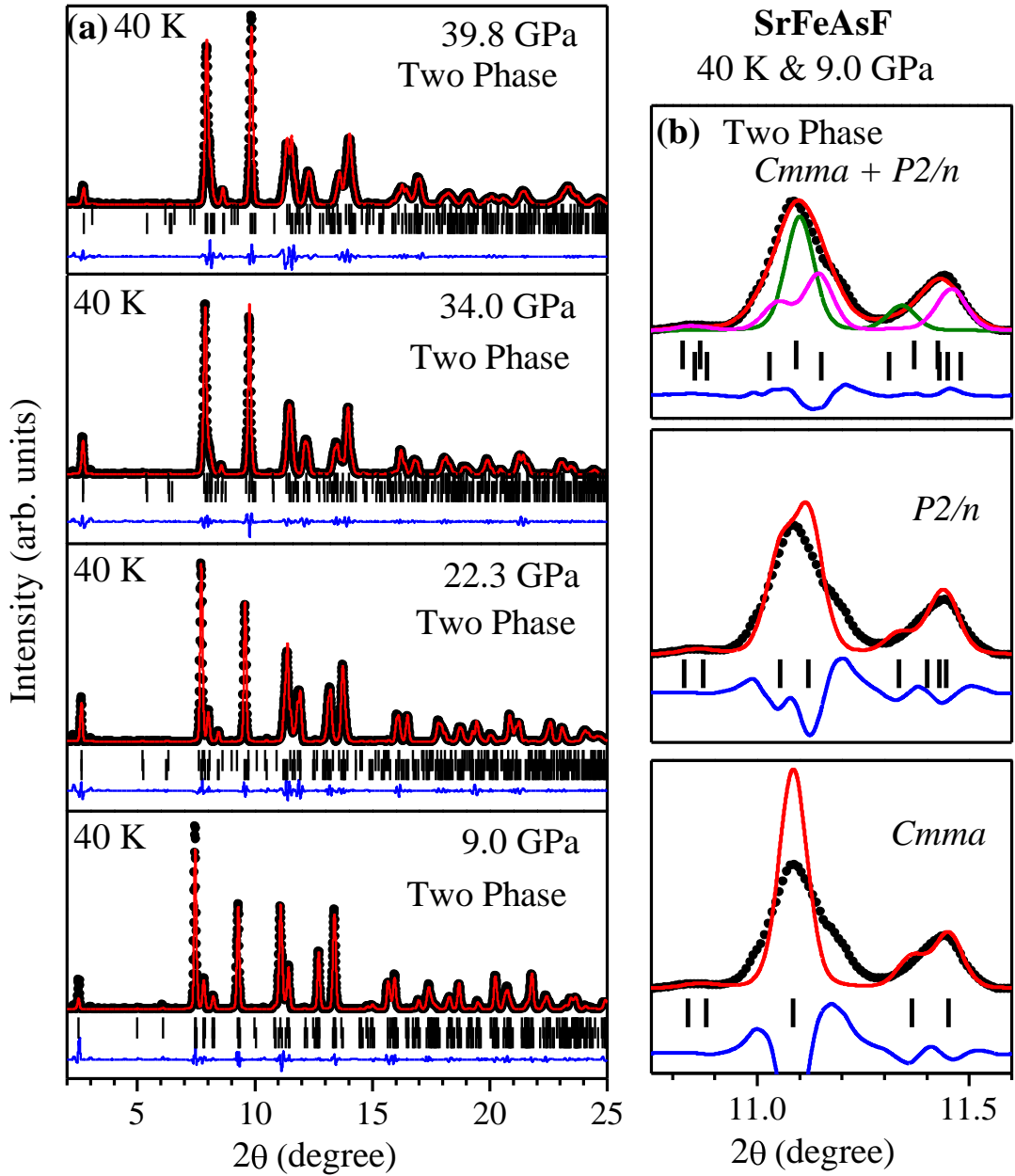

**Figure 6.** (Color online) (a) Observed (solid black circle), calculated (continuous red line), and difference (bottom blue line) profiles obtained after the Rietveld refinement of SrFeAsF at selected pressures and 40 K. The diffraction profiles at 9.0, 22.3, 34.0 and 39.8 GPa are refined using a combination of orthorhombic (*Cmma*) and monoclinic (*P2/n*) phases. Upper and lower vertical tick marks above the difference profiles indicate peak positions of orthorhombic (*Cmma*) and monoclinic (*P2/n*) phases, respectively. (b) The refinement of the diffraction pattern at 9.0 GPa and 40 K with an orthorhombic phase (space group *Cmma*), a monoclinic (*P2/n*) phase, and a combination of orthorhombic (space group *Cmma*) and monoclinic (*P2/n*) phases.

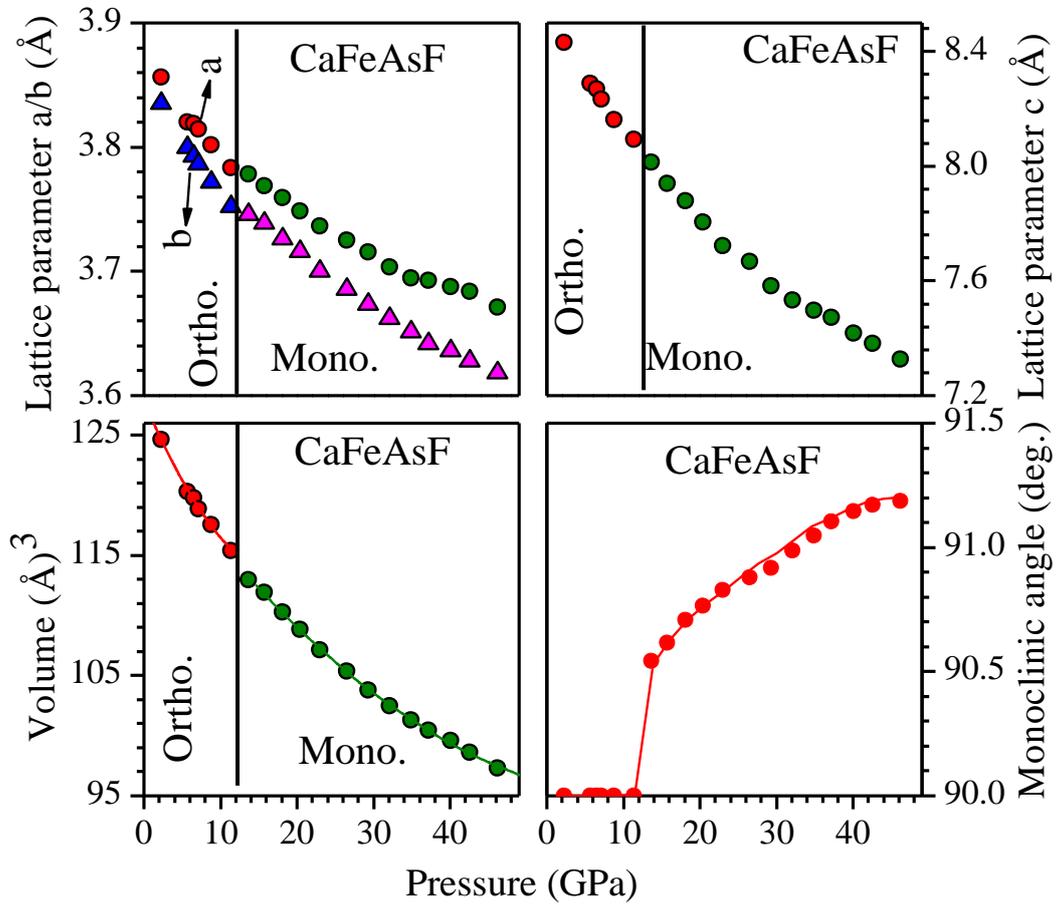

**Figure 7.** (Color online) Pressure dependence of the structural parameters (lattice parameters, volume and monoclinic angle) of CaFeAsF at 40 K as obtained after Rietveld analyses of data with increasing pressure. For sake of comparison, lattice parameters of monoclinic phases are plotted in standard (ab<u>c</u>: *P112/n*) setting.

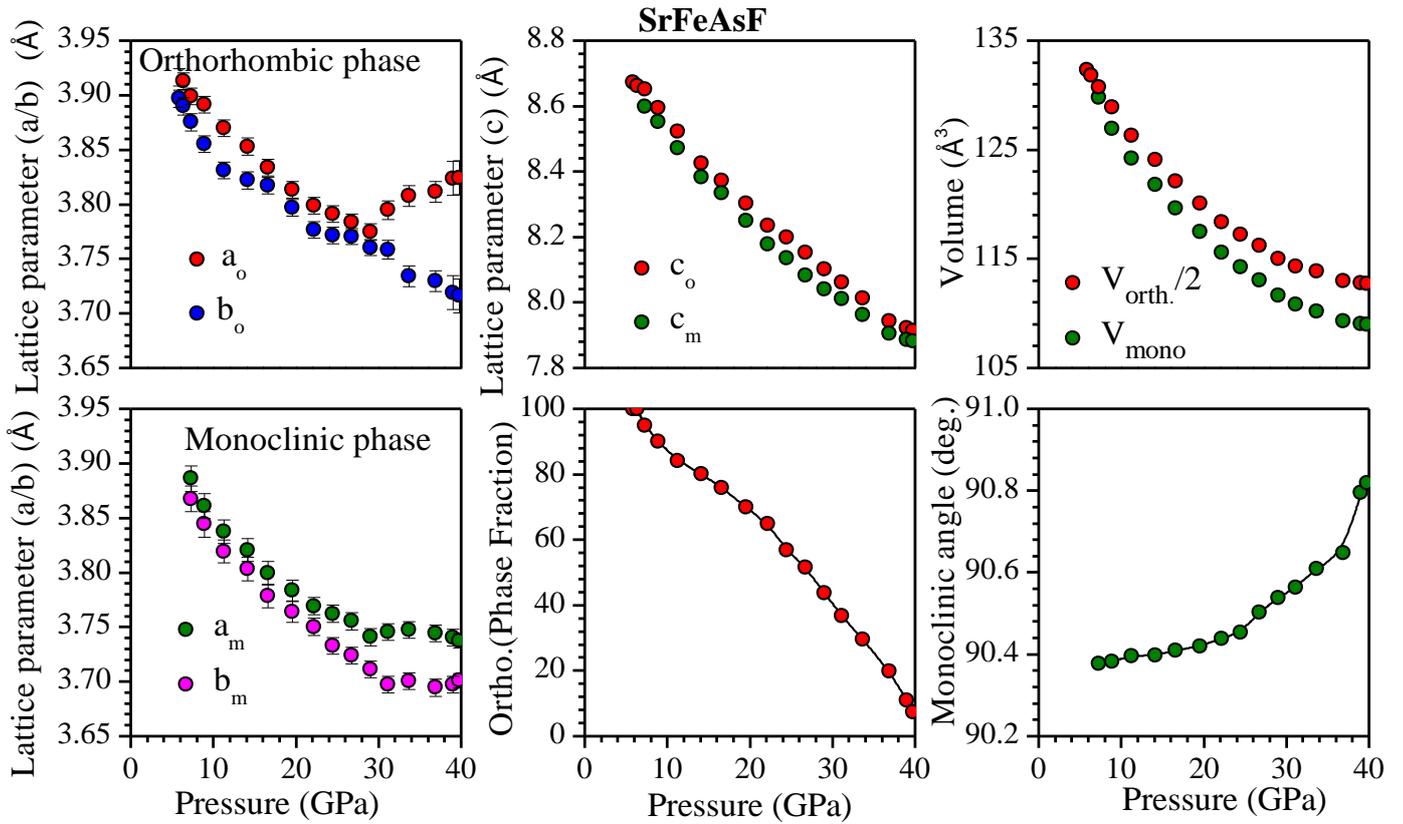

Fig. 8

**Figure 8.** (Color online) Pressure dependence of the structural parameters (lattice parameters, volume, and monoclinic angle) and orthorhombic phase fraction in SrFeAsF at 40 K as obtained after Rietveld analyses of data with increasing pressure. For sake of comparison, lattice parameters along [100], [010] and volume of the orthorhombic phase are divided by $\sqrt{2}$ and 2 respectively.

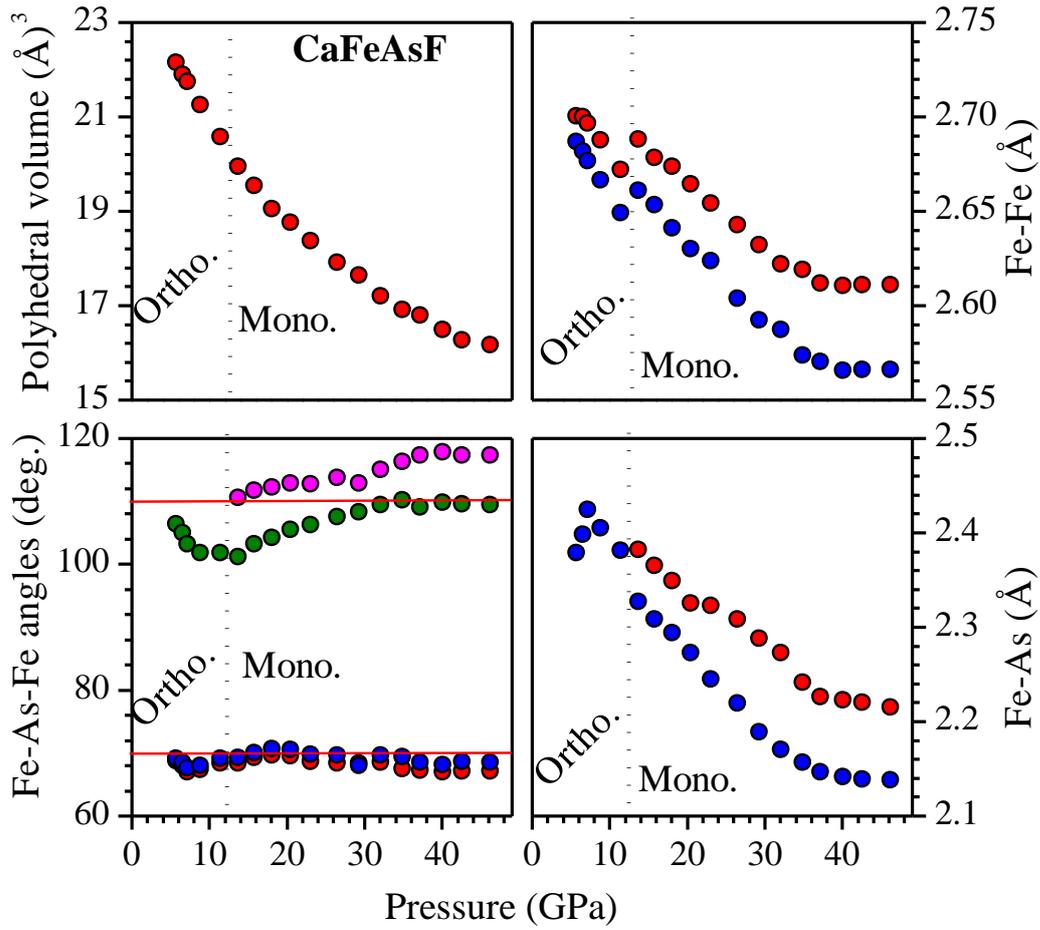

**Figure 9.** (Color online) Pressure dependence of the polyhedral volume, Fe-As-Fe bond angle and Fe-Fe/Fe-As bond length of CaFeAsF at 40 K as obtained after Rietveld analyses of data with increasing pressure.

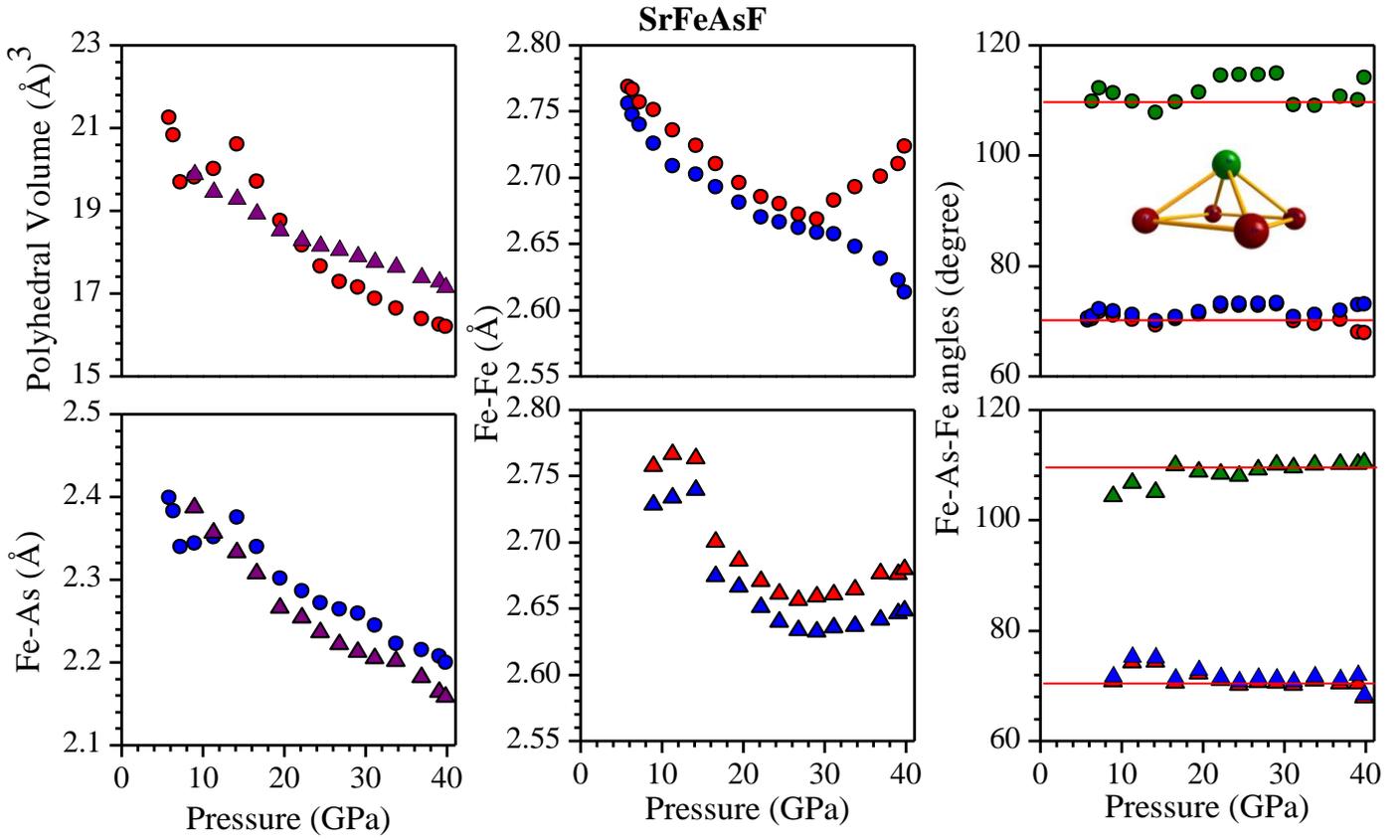

**Figure 10.** (Color online) Pressure dependence of the polyhedral volume, As-Fe-As bond angle and Fe-Fe/Fe-As bond length of SrFeAsF at 40 K as obtained after Rietveld analyses of data with increasing pressure. Solid circle and triangle symbols correspond to the orthorhombic and monoclinic phases respectively.

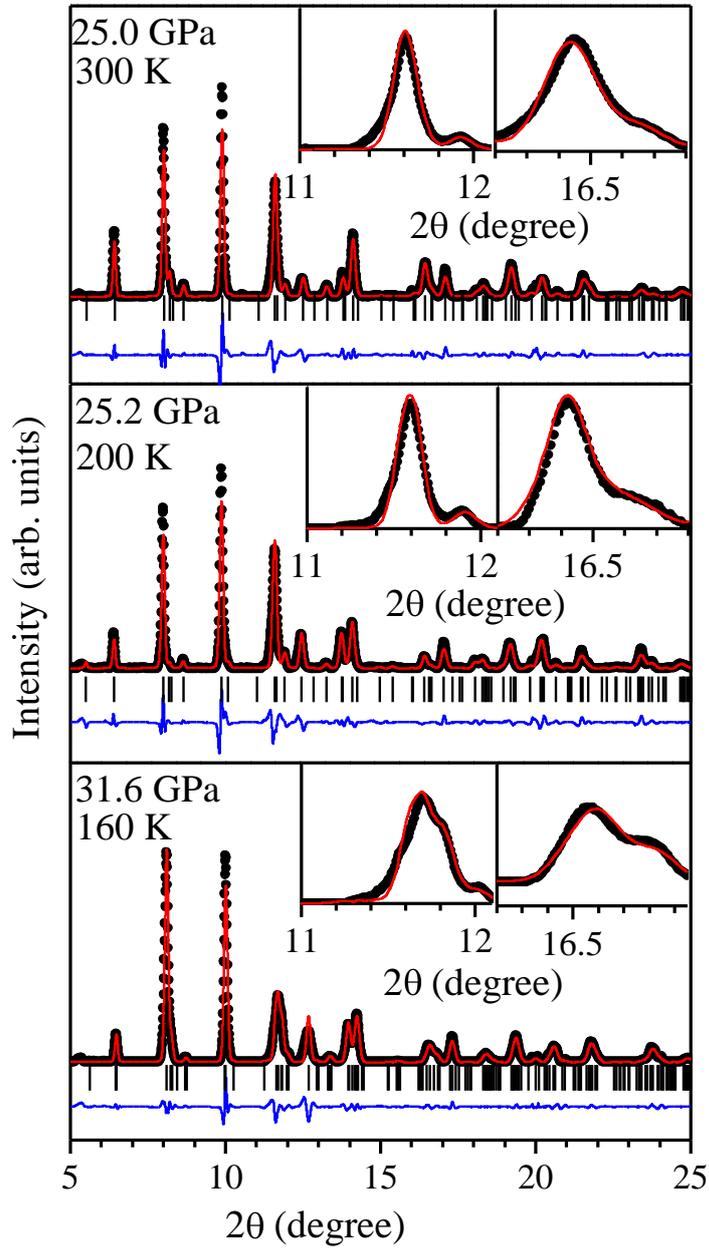

**Figure 11.** (Color online) Observed (solid circle: black), calculated (continuous red line), and difference (bottom blue line) profiles obtained after the Rietveld refinement of CaFeAsF at selected pressures and temperature. The diffraction profiles at 31.6 GPa & 160 K is refined using the monoclinic structure (space group *P2/n*) other refinements at 25.2 GPa & 200 K and 25.0 GPa & 300 K used in tetragonal structure (space group *P4/nmm*). Insets show accountability of certain Bragg's reflections.